\documentclass[aps,pre,reprint,nofootinbib,twocloumn]{revtex4-1}
\usepackage{hyperref}
\usepackage{amsfonts}
\usepackage{amssymb}
\usepackage{amsmath}
\usepackage{graphicx}
\usepackage{epsfig}
\usepackage{color}

\begin{document}

\title{The role of the nature of the noise in the thermal conductance of mechanical systems }
\author{Welles A.~M. \surname{Morgado}}

\affiliation{Department of Physics, PUC-Rio and \\
            National Institute of Science and Technology for Complex Systems \\
            Rua Marqu\^es de S\~ao Vicente 225, G\'avea, CEP 22453-900 RJ, Rio de
            Janeiro, Brazil}
\email{e-mail: welles@fis.puc-rio.br}

\author{S\'{\i}lvio M. \surname{Duarte~Queir\'{o}s}}
\affiliation{Istituto dei Sistemi Complessi, CNR \\
            Via dei Taurini 19, 00185 Roma, Italy}
\email{e-mail: sdqueiro@gmail.com}

\begin{abstract}
Focussing on a paradigmatic small system consisting of two coupled damped
oscillators, we survey the role of the L\'{e}vy-It\^{o} nature of the noise
in the thermal conductance. For white noises, we prove that the L\'{e}vy-It\^{o} composition (Lebesgue measure) of
the noise is irrelevant for the thermal conductance of a non-equilibrium
linearly coupled chain, which signals the independence
between mechanical and thermodynamical properties. On the
other hand, for the non-linearly coupled case, the two types of properties mix
and the explicit definition of the noise plays a central role.
\end{abstract}

\maketitle

\section{Introduction}
The law of heat conduction, or Fourier's law, \emph{i.e.}, the property by which
the heat flux density is equal to the product of the thermal
conductivity by the negative temperature gradient~\cite{huang} is a
paradigmatic manifestation of the ubiquitous laws of thermodynamics~\cite{physics}.
Recently, it has stoked a significant amount of work on its explicit
derivation for large Hamiltonian systems~\cite{lepri,
dhar,kundu}. In this context, models with anharmonic coupling
succeed in diffusing energy, but the analytic
solutions thereto are very demanding, even for the few cases where
that is possible. Since they allow a larger number of
exactly solvable cases, small systems are worthwhile~\cite{dykman} and
particularly relevant in chemical physics and nanosystems~\cite{small-refs}.
In the scope of analytical methods, we highlight the time averaging
of observables that endure a stationary state~\cite{dosp,dosp3}.
This account has several advantages, namely compared with the Fokker-Planck approach,
which cannot be applied to those cases where higher than second order cumulants of the noise
are non-vanishing and also significant. This comprises Poissonian~\cite{poisson,poisson-geral,baule}
and other non-Gaussian massive particles~\cite{welles} as well as  other cases where the interaction
with a reservoir is described by a process with a non-zero singular part of the measure when a L\'{e}vy-It\^{o} (LI)
decomposition is applied~\cite{applebaum}.

Stemming from these facts, we perform a time averaging study
of a small non-equilibrium system composed of two damped
coupled oscillators at distinct temperatures and determine the explicit
formula of the Fourier's law for linear and non-linear cases. In spite of
its simplicity, the former has relevant traits: \emph{i)} it
is a non-equilibrium system; \emph{ii)} Its heat flux definition
is well known; \emph{iii)} It is adjustable to different kinds of reservoirs;
\emph{iv)} It can be expanded into a infinite chain with a nearly
direct application of the results of a $N=2$ block; \emph{v)} It represents the
result of Langevin coloured noises by a renormalisation of the masses \cite%
{dosp} and \emph{vi)} Linearity is still a source of important results in
many areas~\cite{spohn,hanggi12,refrigerator,dhar-refs}.

\section{Model}

Our problem focus on solving the set of equations,
\begin{equation}
m\frac{dv_{i}\left( t\right) }{dt}=-k\,x_{i}\left( t\right) -\gamma
\,v_{i}\left( t\right) -\sum\limits_{l=1}^{2}k_{2\,l-1}\,\left[ x_{i}\left(
t\right) -x_{j}\left( t\right) \right] ^{2\,l-1}+\eta _{i}\left( t\right)
\label{system1}
\end{equation}
with $v_{i}\left( t\right) \equiv \frac{dx_{i}\left( t\right) }{dt}$,
where $(i,j)\in \{1,2\}$ and $k_{1}$ and $k_{3}$ are the linear and
non-linear coupling constants, respectively. The system is decoupled
(linear) for $k_{1\left( 3\right) }=0$. The transfer flux, $j_{12}(t)$, between the two
particles reads,
\begin{equation}
j_{12}\left( t\right) \equiv -\sum\limits_{l=1}^{2}\frac{k_{2\,l-1}}{2}\,%
\left[ x_{1}\left( t\right) -x_{2}\left( t\right) \right] ^{2\,l-1}\left[ v_{%
{1}}\left( t\right) +v_{{2}}\left( t\right) \right] .  \label{transferflux}
\end{equation}

The term, $\eta _{i}\left( t\right) $, represents a general uncorrelated L%
\'{e}vy class stochastic process with cumulants,
\begin{eqnarray}
\left\langle \eta _{i_{1}}\left( t_{1}\right) \,\ldots \,\eta _{i_{n}}\left(
t_{n}\right) \right\rangle _{c} & = & \mathcal{A}(t_{1},n)\,\delta
_{i_{1}\,i_{2}}\,\ldots \delta _{i_{n-1}\,i_{n}} \times \notag \\
 & & \delta \left(
t_{1}-t_{2}\right) \ldots \delta \left( t_{n-1}-t_{n}\right) .  \label{noise}
\end{eqnarray}
From~\cite{marcin}, we either have two or infinite non-zero
cumulants. The former corresponds to the case in which the measure is
absolutely continuous, characterising a Brownian process. In Eq.~(\ref{noise}), $\mathcal{A}(t,n)$ is described by
the noise; If it is Wiener-like, $W(t)\equiv
\int_{t_{0}}^{t}\,\eta (t^{\prime })\,dt^{\prime }$, $\mathcal{A}(t,n)$ is
time-independent and equal to $\sigma ^{2}$ for $n=2$ and zero otherwise ($%
\sigma $ is the standard deviation of the Gaussian). Among infinite non-zero cumulant noises,
we can include the Poisson process for which $
\mathcal{A}(t,n)$ equals $\overline{\Phi ^{n}}\,\lambda (t)$~\cite{hanggi},
with $\Phi $ being the $p(\Phi )$ independent and identically distributed
magnitude and $\lambda (t)$ the rate of shots. Herein, $%
\mathcal{A}$ is time-independent without loss of generality. For $k_{1}=k_{3}=0$, Eq.~(\ref{system1}) is totally decoupled and the
solutions to the problem of homogeneous and sinusoidal heterogeneous Poisson
noises can be found in Ref.~\cite{poisson}.

\section{Results}

Laplace transforming $x _{i}(t)$ and $v _{i}(t)$ we obtain,
\begin{widetext}
\begin{equation}
\begin{array}{c}
\tilde{x}_{i}\left( s\right) =\frac{k_{1}}{R\left( s\right) }\tilde{x}%
_{j}\left( s\right) +\frac{\tilde{\eta}_{i}\left( s\right) }{R\left(
s\right) }+\frac{k_{3}}{R\left( s\right) }\lim_{\alpha \rightarrow 0}\iiint
\frac{\prod\limits_{n=1}^{3}\frac{dq_{n}}{2\,\pi }\left[ \tilde{x}_{i}\left(
\mathrm{i}\,q_{n}+\alpha \right) -\tilde{x}_{j}\left( \mathrm{i}%
\,q_{n}+\alpha \right) \right] }{s-\left( \mathrm{i}\,q_{1}+\mathrm{i}%
\,q_{2}+\mathrm{i}\,q_{3}+3\alpha \right) }  \vspace{0.3cm}\\
s \, \tilde{x}_{i}(s)=\tilde{v}_{i}(s)%
\end{array} ,  \label{laplace1}
\end{equation}
\end{widetext}
(\textrm{Re}$(s)>0$) with $R(s)\,\equiv \left( m\,s^{2}+\gamma
\,s+k+k_{1}\right) $. The solutions to Eq.~(\ref{laplace1}) are
obtained considering the relative position, $\tilde{r}_{D}\left( s\right)
\equiv \tilde{x}_{1}\left( s\right) -\tilde{x}_{2}\left( s\right) $, the
mid-point position, $\tilde{r}_{S}\left( s\right) \equiv \left( \tilde{x}%
_{1}\left( s\right) +\tilde{x}_{2}\left( s\right) \right) /2$, as well as
the respective noises $\tilde{\eta}_{D}\left( s\right) \equiv \tilde{\eta}%
_{1}\left( s\right) -\tilde{\eta}_{2}\left( s\right) $ and $\tilde{\eta}%
_{S}\left( s\right) \equiv \left( \tilde{\eta}_{1}\left( s\right) +\tilde{\eta}%
_{2}\left( s\right) \right) /2$. After some algebra it yields,
\begin{equation}
\left\{
\begin{array}{c}
\tilde{r}_{D}(s)=\frac{\tilde{\eta}_{D}(s)}{R^{\prime }(s)}-\frac{2\,k_{3}}{%
R^{\prime }(s)}\,\underset{\alpha \rightarrow 0}{\lim }\,\int_{-\infty
}^{+\infty }\frac{\prod\nolimits_{l=1}^{3}\frac{dq_{l}}{2\pi }\,\tilde{r}%
_{D}(\mathrm{i}\,q_{l}+\alpha )}{s-\sum\nolimits_{l=1}^{3}\left( \mathrm{i}%
\,q_{l}+\alpha \right) } \\
\tilde{r}_{S}=\frac{\tilde{\eta}_{S}(s)}{R^{\prime \prime }(s)}%
\end{array}%
\right. ,
\label{laplace}
\end{equation}%
with $R^{\prime }(s)\,\equiv \left( m\,s^{2}+\gamma \,s+k+2\,k_{1}\right) $
and $R^{\prime \prime }(s)\,\equiv \left( m\,s^{2}+\gamma \,s+k\right) $.
Reverting Eq.~\ref{laplace} we get $\tilde{x}_{1}\left( s\right) $
and $\tilde{x}_{2}\left( s\right) $. Concomitantly, we must compute the
Laplace transforms of $\eta _{1}$ and $\eta _{2}$,
\begin{equation}
\begin{array}{ccc}
\left\langle \tilde{\eta}_{i_{1}}\left( z_{1}\right) \,\,\ldots \,\tilde{\eta%
}_{i_{n}}\left( z_{n}\right) \right\rangle _{c} & = & \int_{0}^{\infty
}\prod\limits_{j=1}^{n}dt_{i_{j}}\,\exp \left[ -\sum_{j=1}^{n}z_{i_{j}%
\,}t_{i_{j}}\right] \vspace{0.2cm} \\
& & \times \left\langle \eta _{i_{1}}\left( t_{1}\right) \,\,\ldots
\,\eta _{i_{n}}\left( t_{n}\right) \right\rangle _{c} \vspace{0.2cm} \\
& = & \frac{\mathcal{A}(n)}{\sum_{j=1}^{n}z_{i_{j}\,}}\delta
_{i_{1}\,i_{2}}\,\ldots \,\delta _{i_{n-1}\,i_{n}}.%
\end{array}
\label{laplacenoise}
\end{equation}%
that are employed in the averages over time~\cite{dosp},
\begin{widetext}
\begin{eqnarray}
\overline{\left\langle x_{{a}}^{m}\,v_{{b}}^{n}\right\rangle _{c}}
&=&\lim_{z\rightarrow 0}\,z\int \int \int \,\delta \left( t-t_{1}\right)
\,\delta \left( t-t_{2}\right) \,\mathrm{e}^{-z\,t}\left\langle
x_{a}^{m}\left( t_{1}\right) \,v_{b}^{n}\left( t_{2}\right) \right\rangle
_{c}\,dt_{1}\,dt_{2}\,dt\,\,, \\
&=&\lim_{z,\varepsilon \rightarrow 0}\int_{-\infty }^{\infty
}\prod_{l=1}^{m+n}\frac{dq_{l}}{2\pi }z\,\,\frac{\prod_{l^{\prime }=1}^{n}(%
\mathrm{i\,}q_{l^{\prime }}+\varepsilon )\,\left\langle \prod_{l=1}^{m+n}%
\tilde{x}(\mathrm{i\,}q_{l}+\varepsilon )\right\rangle _{c}}{%
z-(\sum_{l^{\prime }=1}^{m+n}\mathrm{i\,}q_{l^{\prime }}+(m+n)\varepsilon )}.
\end{eqnarray}
\end{widetext}
Allowing for a contour that goes along the straight line from $-\rho +%
\mathrm{i}\,\varepsilon $ to $\rho +\mathrm{i}\,\varepsilon $ and then
counterclockwise along a semicircle centered at $0+\mathrm{i}\,\varepsilon $
from $\rho +\mathrm{i}\,\varepsilon $ to $-\rho +\mathrm{i}\,\varepsilon $ ($%
\rho \rightarrow \infty $ and $\varepsilon \rightarrow 0$), we realise that
two situations occur: either the calculation of the residues leads us to a
term like $\frac{z}{z-w}\,u$, with $(u,w)\neq 0$, which vanishes in the
limit $z\rightarrow 0$, or to $\frac{z}{z}u$, which is non-zero.
The problem solving now resumes to an expansion of Eq.~(\ref{transferflux}) in powers of
$k_{3}$, actually the expansion is carried out in powers of \mbox{$k_3 \, T / k_1 ^2$}~\cite{wamm-smdq-extended} (see Appendices).
Hereinafter, besides the results of the Brownian thermostats, which corresponds to the pure continuous measure,
we make explicit the conductance for the Poisson reservoirs, the epitome of singular measures. Nevertheless, the
results for other non-Gaussian noises can be obtained following our methodology yielding the same qualitative results.
In first order the transfer flux reads,
\begin{equation}
\overline{\left\langle j_{12}\right\rangle }=\overline{\left\langle
j_{12}^{\left( 0\right) }\right\rangle }+\overline{\left\langle
j_{12}^{\left( 1\right) }\right\rangle }+\overline{\left\langle
j_{12}^{\left( s\right) }\right\rangle }+\mathcal{O}\left( k_{3}^{2}\right) ,
\label{jtot}
\end{equation}
with,
\begin{equation}
\left\{
\begin{array}{c}
\overline{\left\langle j_{12}^{\left( 0\right) }\right\rangle }=-\frac{%
k_{1}^{2}}{4}\frac{\left[ \mathcal{A}_{1}(2)-\mathcal{A}_{2}(2)\right] }{%
m\,k_{1}^{2}+\gamma ^{2}\left( k+k_{1}\right) } \\
\\
\overline{\left\langle j_{12}^{\left( 1\right) }\right\rangle }=-\frac{3}{8}%
\gamma \,k_{1}\,k_{3}\frac{\left( 2\,k+k_{1}\right) \left[ \mathcal{A}%
_{1}(2)^{2}-\mathcal{A}_{2}(2)^{2}\right] }{\left( k+2\,k_{1}\right) \left[
\gamma ^{2}\left( k+k_{1}\right) +m\,k_{1}^{2}\right] ^{2}}%
\end{array}%
\right. ,
\label{jcont}
\end{equation}
and
\begin{equation}
\overline{\left\langle j_{12}^{\left( s\right) }\right\rangle }=-\frac{27}{2}%
\,{\gamma }^{2}\frac{k_{1}\,k_{{3}}}{\lambda }\frac{\mathcal{N}}{\mathcal{D}}%
\left( \left[ \mathcal{A}_{{1}}(2)^{2}-\mathcal{A}_{{2}}(2)^{2}\right]
\right) ,
\label{jsing}
\end{equation}
for the Poisson case and $\overline{\left\langle j_{12}^{\left( s\right)
}\right\rangle }=0$ for the Brownian case (see Appendices). For Poisson, when $\lambda ^{-1} \ll 1$
(keeping the temperature fixed),
the weight of the singularity of the noise measure dwindles and $\overline{\left\langle j_{12}^{\left( s\right)
}\right\rangle }\rightarrow 0$. The coefficients in Eq.~(\ref{jsing}) are,
\begin{eqnarray}
\mathcal{N} & \equiv & \gamma ^{2}\left( 5\,\,k+3\,k_{1}\right) +m\left(
3\,k_{1}^{2}+4\,k^{2}+11\,k\,k_{1}\right) \, , \vspace{0.5cm}\\
\mathcal{D} &\equiv &  \left[
m\left( 4\,k+9\,k_{1}\right) ^{2}+6\,\gamma ^{2}\left( 2\,k+3\,k_{1}\right) %
\right] \times \\
 & & \left[ 3\,\gamma ^{4}+m^{2}\,k_{1}^{2}+4\,m\,\gamma ^{2}\left(
k+k_{1}\right) \right] \left[ \gamma ^{2}\left( k+k_{1}\right) \right] \notag .
\end{eqnarray}
Thence, we are finally in the position to compute the thermal conductance,%
\begin{eqnarray}
\kappa  &\equiv &-\frac{\partial }{\partial \Delta T}\left\langle
j_{12}\right\rangle _{\Delta T},  \label{fourierlaw} \\
\kappa  &= &-\frac{\overline{\left\langle j_{12}\right\rangle }}{%
T_{1}-T_{2}}.
\notag
\end{eqnarray}%
Resorting to single particle results and the equipartition theorem~\cite{poisson}, we
relate the cumulants of the noise and the proper temperature, $T_{i}$%
, namely, $\mathcal{A}_{i}(2)=2\,\gamma \,T_{i}$, yielding a thermal
conductance, $\kappa =\kappa ^{\left( 0\right) }+\kappa ^{\left( 1\right)
}+\kappa ^{\left( s\right) }+\mathcal{O}\left( k_{3}^{3}\right) $.
Equations (\ref{jtot})-(\ref{jsing}) pave the way to the following assertion;
\emph{When interacting particles are subject to white reservoirs and coupled
in a linear form, the explicit thermal conductance is independent of the
specific nature of the noise, namely the outcome of their L\'{e}vy-It\^{o}
decomposition, whereas for non-linear coupling the nature of the measure of
the noise (its decomposition) is pivotal}. In other words, the linear case is heedless of the
measure of the reservoirs and it only takes into
consideration their temperatures. Hence, $\kappa ^{\left(
0\right) }$ is exactly the same, either we have a Wiener noise (continuous
measure), which is the standard noise in fundamental statistical mechanics
studies \cite{dosp3,li}, or a Poisson noise (paradigmatic case of singular measure). Although
the linear coupling result has only been explicitly proved for two particles, it is
valid for general $N$. In fact, \emph{for a linear chain, the
local energy flow }$\left\langle j_{i,i+1}\right\rangle $\emph{\ can be
written as a function of the cumulants }$\left\langle \eta
_{i}(z)\eta _{1}(z^{\prime })\right\rangle _{c}=2\,\gamma \,T_{i}$\emph{\
and }$\left\langle \eta _{i+1}(z)\eta _{i+1}(z^{\prime })\right\rangle
_{c}=2\,\gamma \,T_{i+1}$\emph{, wherein the dependence on the specific
nature of the noise is eliminated, except for the respective temperatures}.
On the other hand, if the nature of the noise affects the
conductance of the simplest coupling element, the conductance for
generic chains is also changed. This result is unexpected since contrarily to single particle linear cases,
wherein the LI nature is already relevant~\cite{poisson,welles}, for coupled systems the LI composition does solely become significant when the interaction
between the elements of the system happens in a non-linear way. Only in this
case higher-order cumulants of the noise, which can be learnt as higher-order sources of
energy, influence the result. For the same $\left(k_{1},k_{3}\right) $,
in decreasing the singularity by soaring $\lambda $,
the two thermal conductances tally.

\begin{figure}[tbh]
\begin{center}
\includegraphics[width=0.85\columnwidth,angle=0]{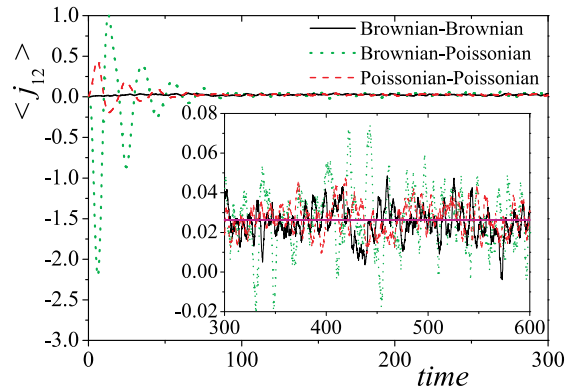}
\end{center}
\caption{(Colour on-line) Average exchange flux $\left\langle j_{12}\right\rangle $ of a two
massive particle system for different combinations of paradigmatic types of noise
with $T_{1}=10$ and $T_{2}=121/10$, $m=10$, $\protect\gamma =k=1$, $%
k_{1}=1/5 $, $k_{3}=0$ and $\protect\lambda =10$ for Poissonian particles.
After the trasient, $\kappa $ agrees with the theoretical value, $\protect\kappa %
=21/800=0.02625 $, with the fitting curves lying within line thickness.
The averages have been obtained averaging over $%
850\times \left( 5\times 10^{5}\right) $ points. The discretisation
used is $\protect\delta t = 10^{-5}$ with snapshots at every $\protect\Delta t = 10^{-3}$.}
\label{conductance}
\end{figure}

To further illustrate these results, we have simulated cases of equally
massive particles subject to Wiener and Poisson noises at different
temperatures, $T$. For the former, we have $T=\sigma ^{2}/2$ whereas for the
latter, we have assumed a homogeneous Poisson process with a rate of events $%
\lambda $, with a random amplitude, $\Phi $, exponentially distributed, $%
p(\Phi )\sim \exp [-\Phi /\bar{\Phi}]$, which yields $T=\lambda _{0}\,\bar{%
\Phi}^{2}/\gamma $~\cite{poisson}. In Fig.~\ref{conductance}, we depict
linear coupling. It is visible that after a transient time, $t^{\ast }$, the
system reaches a stationary state and $\left\langle j_{12}\right\rangle $
becomes equal to $\overline{\left\langle j_{12}\right\rangle }$,
whatever the reservoirs. In fact, even more complex
models, such as linear chains of oscillators, verify the $\kappa =\kappa
^{\left( 0\right) }$ property. Still, this is valid when each particle is
perturbed by different types of noise, \emph{e.g.}, a Brownian particle
coupled with a Poissonian particle. The instance where the noises are of
different nature gives rise to an apparent larger value of the standard
deviation.\footnote{%
Although the computation of $\sigma_{j_{12}}^{2}\equiv \left\langle j_{12}^{2}\right\rangle -\left\langle
j_{12}\right\rangle ^{2}$ is possible, we have set it by as it
demands a mathematical \emph{tour de force} likely to yield a lengthy
formula with little grasping information.}

\begin{figure}[tbh]
\begin{center}
\includegraphics[width=0.85\columnwidth,angle=0]{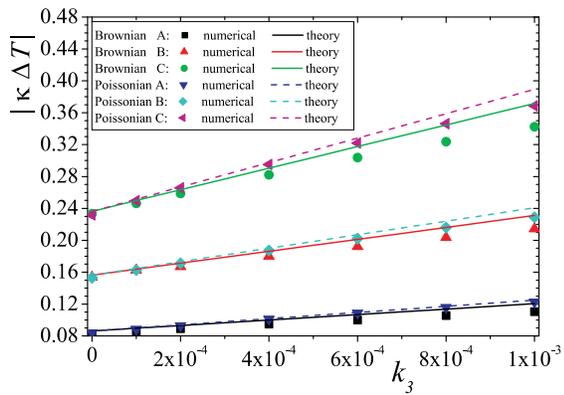}
\end{center}
\caption{(Colour on-line) Comparison between numerically obtained values (symbols) and the
first order approximation of thermal conductance from Eqs.~(\protect\ref%
{jtot})-(\protect\ref{jsing}) for different temperatures pairs,
namely $\mathrm{A}=\left\{10,\frac{169}{10}\right\} $, $\mathrm{B}=\left\{ 10,\frac{225}{10}\right\} $%
, $\mathrm{C}=\left\{ 10,\frac{289}{10}\right\} $ with $m=10$, $\protect%
\gamma =k=1$, $k_{1}=1/5$ and $\protect\lambda =1$ for
Poissonian particles.}
\label{conds}
\end{figure}

In turning $k_{3} \neq 0$, the composition of the measure of the reservoirs
comes into play. In Fig.~\ref{conds}, we show the difference between equivalent
Brownian and Poissonian particles with a good agreement between the averages over
numerical realisations and the respective (first order) approximation. For the same temperature, the larger the value of $k$,
the larger the value of $k _{3} ^{*}$ defining a $10 \%$ difference between numerical values the approximation.

\section{Final Remarks}

To summarise, we have studied the thermal conduction in a paradigmatic
mechanical system composed of two coupled damped harmonic oscillators
subject to generic noises, which can be understood as a concise way to
describe non-equilibrium problems. By averaging in the Laplace space, we
have been able to determine the conductance of a linearly coupled
system and approximate formulae for non-linearly coupled particles. We have
shown the conductance of the former is independent of the nature of
the (white) noise, namely its L\'{e}vy-It\^{o} decomposition structure. This
result is unexpected since the measure of the thermal bath plays a major
role for single particle properties. The dependence on the noise only emerges
when there exists transfer of energy in a non-linear way and higher-order
cumulants of the noise enter in the calculations. In the case of Poisson
noises, we show that the difference to Brownian noises becomes negligeble
when the ratio between the coupling constants and the rate of events is
small.
Our calculations evidence the independence of the thermodynamical properties of the
system from the nature of the reservoirs in linearly coupled systems. On the other hand,
when the coupling is non-linear, the nature of
the reservoirs affects the conductance, which represents a mixture between mechanic and thermodynamical
properties of the system.

Our results have direct implication for the study of the thermal
conductance of systems under the influence of noises other than Wiener, for instance: \emph{i)}
solid state problems wherein shot (singular measure) noise is related to the quantisation of the charge~\cite{nano}; \emph{ii)}
RLC circuits with injection of power at some rate resembling heat pumps~\cite{poisson-geral}; \emph{iii)} Surface diffusion and low vibrational motion with adsorbates, \emph{e.g.},  Na/Cu(001) compounds~\cite{martinez-casado}; \emph{iv)} Biological motors in which shot noise mimics the nonequilibrium stochastic hydrolysis of adenosine triphosphate~\cite{baule} and \emph{v)} Molecular dynamics when the Andersen thermostat is applied. Actually, in molecular dynamics~\cite{baths}, the Langevin reservoir is just one in a large collection of baths represented by our definition of noise~(\ref{noise}). In these problems, for non-linearly coupled elements the experimentally measured energy flux will be greater than the energy flux given by Langevin reservoirs at the same temperatures and equal if coupling is linear.
At the theoretical level, the method is worth being used to shed light
on non-linear chains as well. Within this context, the feasible approach
is once again to consider a perturbative expansion of the non-linearities in the problem.

\medskip
\begin{acknowledgments}
We would like to thank the partial funding from Faperj and CNPq (W.A.M.M.) and the European Commission through the Marie Curie Actions FP7-PEOPLE-2009-IEF
(contract nr 250589) (S.M.D.Q.). Diogo O. Soares-Pinto is acknowledged for early discussions on this subject.
\end{acknowledgments}

\begin{widetext}
\appendix
\section{Heat flux definition}

As indicated in the main document the model consists of two masses connected
by linear and non-linear springs so that,%
\begin{equation}
\begin{array}{ccc}
\dot{x}_{1} & = & v_{1}, \\
\dot{x}_{2} & = & v_{2}, \\
m\,\ddot{x}_{1} & = & -k\,x_{1}-k_{1}(x_{1}-x_{2})-k_{3}(x_{1}-x_{2})^{3}-%
\gamma \dot{x}_{1}+\eta _{1}, \\
m\,\ddot{x}_{2} & = & -k\,x_{2}-k_{1}(x_{2}-x_{1})-k_{3}(x_{2}-x_{1})^{3}-%
\gamma \dot{x}_{2}+\eta _{2}.%
\end{array}%
\end{equation}

\bigskip The flux of energy $J_{1\rightarrow 2}$ defines the thermal
conductance. For this unidimensional mechanical system, the transmitted
power is by the the instantaneous power difference~\cite{dosp3}%
:
\begin{equation}
J_{1\rightarrow 2}=\frac{dW_{1\rightarrow 2}-dW_{2\rightarrow 1}}{2\,dt}=%
\frac{F_{1\rightarrow 2}v_{2}-F_{2\rightarrow 1}v_{1}}{2},
\end{equation}%
where $F_{\alpha \rightarrow \beta }$ is the force exerted on particle $%
\beta $ by particle $\alpha $.

Assuming a conservative potential, the form of the force between the
particles will be,%
\begin{equation}
F_{\alpha \rightarrow \beta }=-k_{1}(x_{\alpha }-x_{\beta })-k_{3}(x_{\alpha
}-x_{\beta })^{3}.
\end{equation}

Our goal is to derive a systematic expansion for the time-average $%
j_{12}\equiv \left\langle J_{1\rightarrow 2}\right\rangle \equiv \kappa
\,(T_{2}-T_{1})$, where $\kappa $ is the thermal conductance of the model.
Thus, we can write
\begin{equation}
\begin{array}{ccc}
j_{12} & = & \left\langle \frac{F_{1\rightarrow 2}v_{2}-F_{2\rightarrow
1}v_{1}}{2}\right\rangle \\
& = & -k_{1}\left\langle \frac{x_{1}v_{2}-x_{2}v_{1}}{2}\right\rangle
-k_{3}\left\langle \frac{(x_{1}-x_{2})^{3}(v_{2}+v_{1})}{2}\right\rangle ,%
\end{array}%
\end{equation}%
where the terms of the form $<x_{1}^{n}v_{1}>$ and $<x_{2}^{n}v_{2}>$ vanish
identically as $t\rightarrow \infty $~\cite{dosp3}.

\subsection{The Gaussian case}

The noise functions are assumed to be white and Gaussian,%
\begin{equation}
\begin{array}{ccc}
<\eta _{1}(t_{1})>_{c} & = & 0, \\
<\eta _{1}(t_{1})\eta _{1}(t_{2})>_{c} & = & 2\,\gamma \,T_{1}\,\delta
(t_{1}-t_{2}), \\
<\eta _{2}(t_{1})>_{c} & = & 0, \\
<\eta _{2}(t_{1})\eta _{2}(t_{2})>_{c} & = & 2\,\gamma \,T_{2}\,\delta
(t_{1}-t_{2}),%
\end{array}%
\end{equation}%
The initial conditions will be assumed to be%
\begin{equation}
x_{1}(0)=v_{1}(0)=x_{2}(0)=v_{2}(0)=0.
\end{equation}%
The Laplace transforms read,%
\begin{equation}
\begin{array}{c}
\tilde{v}_{1}(s)=s\tilde{x}_{1}(s), \\
\tilde{v}_{2}(s)=s\tilde{x}_{2}(s),%
\end{array}%
\end{equation}%
using $R(s)\equiv \left( m\,s^{2}\,+\gamma \,s\,+(k+k_{1})\right)
=m(s^{2}+\theta s+\omega ^{2})=(s-\zeta _{+})(s-\zeta _{-})$, where
\begin{equation}
\zeta _{\pm }=-\frac{\theta }{2}\pm \frac{i}{2}\sqrt{4\omega ^{2}-\theta ^{2}%
}
\end{equation}%
we have,%
\begin{equation}
\begin{array}{ccc}
\tilde{x}_{1}(s) & = & \frac{k_{1}}{R(s)}\,\tilde{x}_{2}(s)\,+\frac{\tilde{%
\eta}_{1}(s)}{R(s)}-\frac{k_{3}}{R(s)}\,\lim_{\alpha \rightarrow
0}\int_{-\infty }^{+\infty }\frac{dq_{1}}{2\pi }\,\int_{-\infty }^{+\infty }%
\frac{dq_{2}}{2\pi }\,\int_{-\infty }^{+\infty }\frac{dq_{3}}{2\pi }\,\times
\\
&  &  \\
&  & \times \frac{\left( \tilde{x}_{1}(iq_{1}+\alpha )-\tilde{x}%
_{2}(iq_{1}+\alpha ))\,(\tilde{x}_{1}(iq_{2}+\alpha )-\tilde{x}%
_{2}(iq_{2}+\alpha ))\,(\tilde{x}_{1}(iq_{3}+\alpha )-\tilde{x}%
_{2}(iq_{3}+\alpha )\right) }{s-(iq_{1}+iq_{2}+iq_{3}+3\alpha )},%
\end{array}%
\end{equation}%
and%
\begin{equation}
\begin{array}{ccc}
\tilde{x}_{2}(s) & = & \frac{k_{1}}{R(s)}\,\tilde{x}_{1}(s)\,+\frac{\tilde{%
\eta}_{2}(s)}{R(s)}+\frac{k_{3}}{R(s)}\,\lim_{\alpha \rightarrow
0}\int_{-\infty }^{+\infty }\frac{dq_{1}}{2\pi }\,\int_{-\infty }^{+\infty }%
\frac{dq_{2}}{2\pi }\,\int_{-\infty }^{+\infty }\frac{dq_{3}}{2\pi }\,\times
\\
&  &  \\
&  & \times \frac{\left( \tilde{x}_{1}(iq_{1}+\alpha )-\tilde{x}%
_{2}(iq_{1}+\alpha ))\,(\tilde{x}_{1}(iq_{2}+\alpha )-\tilde{x}%
_{2}(iq_{2}+\alpha ))\,(\tilde{x}_{1}(iq_{3}+\alpha )-\tilde{x}%
_{2}(iq_{3}+\alpha )\right) }{s-(iq_{1}+iq_{2}+iq_{3}+3\alpha )}.%
\end{array}%
\end{equation}%
An straightforward way to write a recurrence equation for the problem above
is to take the difference:%
\begin{equation}
\begin{array}{ccc}
\left( 1+\frac{k_{1}}{\,R(s)}\right) (\tilde{x}_{1}(s)-\tilde{x}_{2}(s)) & =
& \frac{\tilde{\eta}_{1}(s)-\tilde{\eta}_{2}(s)}{R(s)}-\frac{2\,k_{3}}{R(s)}%
\,\lim_{\alpha \rightarrow 0}\int_{-\infty }^{+\infty }\frac{dq_{1}}{2\pi }%
\,\int_{-\infty }^{+\infty }\frac{dq_{2}}{2\pi }\,\int_{-\infty }^{+\infty }%
\frac{dq_{3}}{2\pi }\,\times \\
&  &  \\
&  & \times \frac{\left( \tilde{x}_{1}(iq_{1}+\alpha )-\tilde{x}%
_{2}(iq_{1}+\alpha ))\,(\tilde{x}_{1}(iq_{2}+\alpha )-\tilde{x}%
_{2}(iq_{2}+\alpha ))\,(\tilde{x}_{1}(iq_{3}+\alpha )-\tilde{x}%
_{2}(iq_{3}+\alpha )\right) }{s-(iq_{1}+iq_{2}+iq_{3}+3\alpha )}.%
\end{array}%
\end{equation}%
Defining $R^{\prime }(s)\equiv \left( m\,s^{2}\,+\gamma
\,s\,+(k+2\,k_{1})\right) =(s^{2}+\theta s+\omega _{1}^{2})=(s-\zeta
_{1+})(s-\zeta _{1-})$, where,
\begin{equation*}
\zeta _{1\pm }=-\frac{\theta }{2}\pm \frac{i}{2}\sqrt{4\omega
_{1}^{2}-\theta ^{2}}
\end{equation*}%
we obtain the difference,%
\begin{equation}
\begin{array}{ccc}
\tilde{r}_{D} & = & \tilde{x}_{1}(s)-\tilde{x}_{2}(s) \\
&  &  \\
& = & \frac{\tilde{\eta}_{1}(s)-\tilde{\eta}_{2}(s)}{R^{\prime }(s)}-\frac{%
2\,k_{3}}{R^{\prime }(s)}\,\lim_{\alpha \rightarrow 0}\int_{-\infty
}^{+\infty }\frac{dq_{1}}{2\pi }\,\int_{-\infty }^{+\infty }\frac{dq_{2}}{%
2\pi }\,\int_{-\infty }^{+\infty }\frac{dq_{3}}{2\pi }\,\times \\
&  &  \\
&  & \times \frac{\left( \tilde{x}_{1}(iq_{1}+\alpha )-\tilde{x}%
_{2}(iq_{1}+\alpha ))\,(\tilde{x}_{1}(iq_{2}+\alpha )-\tilde{x}%
_{2}(iq_{2}+\alpha ))\,(\tilde{x}_{1}(iq_{3}+\alpha )-\tilde{x}%
_{2}(iq_{3}+\alpha )\right) }{s-(iq_{1}+iq_{2}+iq_{3}+3\alpha )}.%
\end{array}%
\end{equation}%
Defining $R^{\prime \prime }(s)\equiv \left( m\,s^{2}\,+\gamma
\,s\,+k\right) =(s^{2}+\theta s+\omega _{2}^{2})=(s-\zeta _{2+})(s-\zeta
_{2-})$, where,%
\begin{equation*}
\zeta _{2\pm }=-\frac{\theta }{2}\pm \frac{i}{2}\sqrt{4\omega
_{2}^{2}-\theta ^{2}}
\end{equation*}%
we obtain the sum as,
\begin{equation}
\begin{array}{ccc}
\tilde{r}_{S} & = & \frac{\tilde{x}_{1}(s)+\tilde{x}_{2}(s)}{2} \\
&  &  \\
& = & \frac{\tilde{\eta}_{1}(s)+\tilde{\eta}_{2}(s)}{2\,\,R^{\prime \prime
}(s)}.%
\end{array}%
\end{equation}
Inverting the relations it yields,%
\begin{equation}
\begin{array}{ccc}
\tilde{x}_{1}(s) & = & \tilde{r}_{S}+\frac{\tilde{r}_{D}}{2}, \\
&  &  \\
\tilde{x}_{2}(s) & = & \tilde{r}_{S}-\frac{\tilde{r}_{D}}{2}.%
\end{array}%
\end{equation}%
In the same way, we can define the difference and the average of the noise
as,%
\begin{equation}
\begin{array}{ccc}
\tilde{\eta}_{S}(s) & = & \frac{\tilde{\eta}_{1}+\tilde{\eta}_{2}}{2}, \\
&  &  \\
\tilde{\eta}_{D}(s) & = & \tilde{\eta}_{1}-\tilde{\eta}_{2}.%
\end{array}%
\end{equation}
We can now express the recurrence relations for the new variables,%
\begin{equation}
\begin{array}{ccc}
\tilde{r}_{D}(s) & = & \frac{\tilde{\eta}_{D}(s)}{R^{\prime }(s)}-\frac{%
2\,k_{3}}{R^{\prime }(s)}\,\lim_{\alpha \rightarrow 0}\int_{-\infty
}^{+\infty }\frac{dq_{1}}{2\pi }\,\int_{-\infty }^{+\infty }\frac{dq_{2}}{%
2\pi }\,\int_{-\infty }^{+\infty }\frac{dq_{3}}{2\pi }\,\times \\
&  &  \\
&  & \times \frac{\tilde{r}_{D}(iq_{1}+\alpha )\,\tilde{r}_{D}(iq_{2}+\alpha
)\,\tilde{r}_{D}(iq_{3}+\alpha )}{s-(iq_{1}+iq_{2}+iq_{3}+3\alpha )}, \\
&  &  \\
\tilde{r}_{S} & = & \frac{\tilde{\eta}_{S}(s)}{R^{\prime \prime }(s)}.%
\end{array}%
\end{equation}
The Laplace transforms of the noise are,%
\begin{equation}
\begin{array}{c}
<\tilde{\eta}_{D}(s_{1})\tilde{\eta}_{D}(s_{2})>_{c}=\frac{2\,\gamma
\,(T_{1}+T_{2})}{s_{1}+s_{2}}, \\
<\tilde{\eta}_{S}(s_{1})\tilde{\eta}_{S}(s_{2})>_{c}=\frac{\gamma
\,(T_{1}+T_{2})}{2\,(s_{1}+s_{2})}, \\
<\tilde{\eta}_{S}(s_{1})\tilde{\eta}_{D}(s_{2})>_{c}=\frac{\gamma
\,(T_{1}-T_{2})}{s_{1}+s_{2}}.%
\end{array}%
\end{equation}

\section{Heat conductance}

\subsection{General expression}

The series expansion for the thermal flux reads,
\begin{equation}
j_{12}=-\frac{k_{1}}{2}\left\langle (x_{{1}}-x_{{2}})(v_{{2}}+v_{{1}%
})\right\rangle -\frac{k_{{3}}}{2}\left\langle (x_{{1}}-x_{{2}})^{3}(v_{{2}%
}+v_{{1}})\right\rangle
\end{equation}

We can write ($a,b=1,2$),%
\begin{equation}
\begin{array}{ccc}
\overline{\left\langle x_{{a}}^{m}\,v_{{b}}^{n}\right\rangle _{c}} & = &
\lim_{z\rightarrow 0}\,\lim_{\epsilon \rightarrow 0}\,\int_{-\infty
}^{\infty }\prod_{k=1}^{m}\frac{dq_{k}}{2\pi }\,\int_{-\infty }^{\infty
}\prod_{l=1}^{n}\frac{dq_{l}}{2\pi }\,\frac{z}{z-(\sum_{k=1}^{m}iq_{k}+%
\sum_{l=1}^{n}iq_{l}+(m+n)\epsilon )}\,\,\left\langle \prod_{k=1}^{m}\tilde{x%
}(iq_{k}+\epsilon )\,\prod_{l=1}^{n}\tilde{v}(iq_{l}+\epsilon )\right\rangle
_{c} \\
&  &  \\
& = & \lim_{z\rightarrow 0}\,\lim_{\epsilon \rightarrow 0}\,\int_{-\infty
}^{\infty }\prod_{k=1}^{m+n}\frac{dq_{k}}{2\pi }\,\,\frac{z}{%
z-(\sum_{k=1}^{m+n}iq_{k}+(m+n)\epsilon )}\,\prod_{l=1}^{n}(iq_{l}+\epsilon
)\,\left\langle \prod_{k=1}^{m+n}\tilde{x}(iq_{k}+\epsilon )\right\rangle
_{c}.%
\end{array}%
\end{equation}

We can thus express the heat flux $\overline{\left\langle
j_{12}\right\rangle }$ as,%
\begin{equation}
\begin{array}{ccc}
\overline{\left\langle j_{12}\right\rangle } & = & -k_{1}\lim_{z\rightarrow
0}\,\lim_{\epsilon \rightarrow 0}\,\int_{-\infty }^{\infty }\frac{dq_{1}}{%
2\pi }\,\int_{-\infty }^{\infty }\frac{dq_{2}}{2\pi }\,\,\frac{z}{%
z-(iq_{1}+iq_{2}+2\epsilon )}(iq_{2}+\epsilon )\,\left\langle \tilde{r}%
_{D}(iq_{1}+\epsilon )\,\tilde{r}_{S}(iq_{2}+\epsilon )\right\rangle _{c}-
\\
&  &  \\
&  & -k_{3}\,\lim_{z\rightarrow 0}\,\lim_{\epsilon \rightarrow
0}\,\int_{-\infty }^{\infty }\frac{dq_{1}}{2\pi }\,\int_{-\infty }^{\infty }%
\frac{dq_{2}}{2\pi }\,\int_{-\infty }^{\infty }\frac{dq_{3}}{2\pi }%
\,\int_{-\infty }^{\infty }\frac{dq_{4}}{2\pi }\,\,\frac{z}{%
z-(iq_{1}+iq_{2}+iq_{3}+iq_{4}+2\epsilon )}\,\,\times \\
&  &  \\
&  & \times \left\{ (iq_{4}+\epsilon )\,\left\langle \tilde{r}%
_{D}(iq_{1}+\epsilon )\,\tilde{r}_{D}(iq_{2}+\epsilon )\,\tilde{r}%
_{D}(iq_{3}+\epsilon )\,\tilde{r}_{S}(iq_{4}+\epsilon )\right\rangle
_{c}\right\} .%
\end{array}%
\end{equation}

We now proceed to expand the heat flow in powers of $k_3$.

\subsection{Order zero on $k_3$}

The order zero term is,%
\begin{equation*}
\begin{array}{ccc}
\overline{\left\langle j_{12}\right\rangle _{0}} & = & -k_{1}\lim_{z%
\rightarrow 0}\,\lim_{\epsilon \rightarrow 0}\,\int_{-\infty }^{\infty }%
\frac{dq_{1}}{2\pi }\,\int_{-\infty }^{\infty }\frac{dq_{2}}{2\pi }\frac{%
z\,(iq_{2}+\epsilon )}{z-(iq_{1}+iq_{2}+2\epsilon )}\,\frac{\left\langle
\tilde{\eta}_{D}(iq_{1}+\epsilon )\,\tilde{\eta}_{S}(iq_{2}+\epsilon
)\right\rangle _{c}}{\,R^{\prime }(iq_{1}+\epsilon )\,R^{\prime \prime
}(iq_{2}+\epsilon )} \\
&  &  \\
& = & -\frac{k_{1}}{m^{2}}\lim_{z\rightarrow 0}\,\lim_{\epsilon \rightarrow
0}\,\int_{-\infty }^{\infty }\frac{dq_{1}}{2\pi }\,\int_{-\infty }^{\infty }%
\frac{dq_{2}}{2\pi }\frac{z\,(iq_{2}+\epsilon )\,}{z-(iq_{1}+iq_{2}+2%
\epsilon )}\frac{\left\langle \tilde{\eta}_{D}(iq_{1}+\epsilon )\,\tilde{\eta%
}_{S}(iq_{2}+\epsilon )\right\rangle _{c}}{R^{\prime }(iq_{1}+\epsilon
)\,R^{\prime \prime }(iq_{2}+\epsilon )} \\
&  &  \\
& = & -\frac{\gamma \,(T_{1}-T_{2})\,k_{1}}{m^{2}}\lim_{z\rightarrow
0}\,\lim_{\epsilon \rightarrow 0}\,\int_{-\infty }^{\infty }\frac{dq_{1}}{%
2\pi }\,\int_{-\infty }^{\infty }\frac{dq_{2}}{2\pi }\frac{z}{%
z-(iq_{1}+iq_{2}+2\epsilon )}\frac{(iq_{2}+\epsilon )\,}{R^{\prime
}(iq_{1}+\epsilon )\,R^{\prime \prime }(iq_{2}+\epsilon )}\frac{1}{%
(iq_{1}+iq_{2}+2\epsilon )} \\
&  &  \\
& = & -\frac{1}{2}\,{\frac{\gamma \,\left( T_{{1}}-T_{{2}}\right) k_{1}^{2}}{%
{\gamma }^{2}\left( k+k_{1}\right) +{k}_{1}^{2}m}},%
\end{array}%
\end{equation*}%
which is already a well known result~\cite{dosp3,poisson}.

\subsection{Order 1 on $k_3$}

There are two contributions to order 1 on $k_{3}$: one from the quadratic
term and another from the quartic one. The quadratic term reads,%
\begin{equation}
\begin{array}{ccc}
\overline{\left\langle j_{12}\right\rangle }_{12} & = & k_{1}\lim_{z%
\rightarrow 0}\,\lim_{\epsilon \rightarrow 0}\,\lim_{\alpha \rightarrow
0}\int_{-\infty }^{\infty }\frac{dq_{1}}{2\pi }\,\,\ldots \,\int_{-\infty
}^{+\infty }\frac{dq_{5}}{2\pi }\,\frac{z\,(iq_{2}+\epsilon )\,}{%
z-(iq_{1}+iq_{2}+2\epsilon )}\frac{2\,k_{3}}{R^{\prime }(iq_{1}+\epsilon )}%
\frac{\left\langle \tilde{r}_{D}(iq_{3}+\alpha )\,\tilde{r}%
_{D}(iq_{4}+\alpha )\,\tilde{r}_{D}(iq_{5}+\alpha )\,\tilde{r}%
_{S}(iq_{2}+\epsilon )\right\rangle _{c}}{iq_{1}+\epsilon
-(iq_{3}+iq_{4}+iq_{5}+3\alpha )} \\
&  &  \\
& = & 12\,\gamma ^{2}\,k_{1}\,k_{3}\lim_{z\rightarrow 0}\,\lim_{\epsilon
\rightarrow 0}\,\lim_{\alpha \rightarrow 0}\,\int_{-\infty }^{\infty }\frac{%
dq_{1}}{2\pi }\,\ldots \,\int_{-\infty }^{+\infty }\frac{dq_{5}}{2\pi }\frac{%
z}{z-(iq_{1}+iq_{2}+2\epsilon )}\,\frac{(iq_{2}+\epsilon )}{iq_{1}+\epsilon
-(iq_{3}+iq_{4}+iq_{5}+3\alpha )}\times \\
&  &  \\
&  & \times \frac{1}{R^{\prime }(iq_{1}+\epsilon )\,R^{\prime
}(iq_{3}+\alpha )\,R^{\prime }(iq_{4}+\alpha )\,R^{\prime }(iq_{5}+\alpha
)\,R^{\prime \prime }(iq_{2}+\epsilon )}\frac{(T_{1}+T_{2})\,(T_{1}-T_{2})}{%
(iq_{3}+iq_{4}+2\alpha )(iq_{5}+\alpha +iq_{2}+\epsilon )} \\
&  &  \\
& = & -\frac{3}{2}\gamma \,k_{{3}}{\frac{\left( {\gamma }^{2}k-m{k}%
_{1}^{2}\right) \left( {T_{{1}}}^{2}-{T_{{2}}}^{2}\right) }{\left(
k+2\,k_{1}\right) \left[ m{k}^{2}+{\gamma }^{2}\left( k+k_{1}\right) \right]
^{2}}},%
\end{array}%
\end{equation}

The quartic term reads,
\begin{equation}
\begin{array}{ccc}
\overline{\left\langle j_{12}\right\rangle _{12}} & = & -k_{3}\,\lim_{z%
\rightarrow 0}\,\lim_{\epsilon \rightarrow 0}\,\int_{-\infty }^{\infty }%
\frac{dq_{1}}{2\pi }\,\,\ldots \int_{-\infty }^{\infty }\frac{dq_{4}}{2\pi }%
\frac{z}{z-(iq_{1}+iq_{2}+iq_{3}+iq_{4}+2\epsilon )}\times  \\
&  &  \\
& = & \times \left\{ (iq_{2}+\epsilon )\,\left\langle \tilde{r}%
_{D}(iq_{1}+\epsilon )\,\tilde{r}_{S}(iq_{2}+\epsilon )\,\tilde{r}%
_{D}(iq_{3}+\epsilon )\,\tilde{r}_{D}(iq_{4}+\epsilon )\right\rangle
_{c}\right\}  \\
&  &  \\
& = & -6\,\gamma ^{2}\,k_{3}\,\lim_{z\rightarrow 0}\,\lim_{\epsilon
\rightarrow 0}\,\int_{-\infty }^{\infty }\frac{dq_{1}}{2\pi }\,\,\ldots
\int_{-\infty }^{\infty }\frac{dq_{4}}{2\pi }\frac{z}{%
z-(iq_{1}+iq_{2}+iq_{3}+iq_{4}+2\epsilon )}\times  \\
&  &  \\
&  & \times \frac{(iq_{2}+\epsilon )}{R^{\prime }(iq_{1}+\epsilon
)\,R^{\prime }(iq_{3}+\alpha )\,R^{\prime }(iq_{1}+\epsilon )\,R^{\prime
\prime }(iq_{2}+\epsilon )}\frac{(T_{1}+T_{2})\,(T_{1}-T_{2})}{%
(iq_{3}+iq_{4}+2\alpha )(iq_{1}+iq_{2}+2\epsilon )} \\
&  &  \\
& = & -\frac{3}{2}\,{\frac{\gamma \,k_{1}\,k_{{3}}\left( {T_{{1}}}^{2}-{T_{{2%
}}}^{2}\right) }{\left( k+2\,k_{1}\right) \,\left[ {\gamma }^{2}\left(
k+k_{1}\right) +m\,{k}_{1}^{2}\right] }}.%
\end{array}%
\end{equation}

The total term reads,
\begin{equation}
\overline{\left\langle j_{12}^{\left( 1\right) }\right\rangle }=-\frac{3}{2}%
\,{\frac{{\gamma }^{3}k_{1}\,k_{{3}}\left( 2\,k+k_{1}\right) \left( {T_{{1}}}%
^{2}-{T_{{2}}}^{2}\right) }{\left( k+2\,k_{1}\right) \left[ m{k}^{2}+{\gamma
}^{2}\left( k+k_{1}\right) \right] ^{2}}}.
\end{equation}

By the relation $j_{12}=-\kappa \Delta T$, we obtain the expansion of the
thermal conductance as,
\begin{equation}
\begin{array}{ccc}
\kappa  & = & {\frac{\gamma k_{1}^{2}}{{\gamma }^{2}\left( k+k_{1}\right) +{k%
}_{1}^{2}m}}+\frac{3}{2}\,{\frac{{\gamma }^{3}k_{1}\,k_{{3}}\left(
2\,k+k_{1}\right) \left( {T_{{1}}+T_{{2}}}\right) }{\left( k+2\,k_{1}\right) %
\left[ m{k}^{2}+{\gamma }^{2}\left( k+k_{1}\right) \right] ^{2}}}+\mathcal{O}%
(k_{3}^{2}) \\
&  &  \\
\kappa  & = & \kappa ^{\left( 0\right) }+\kappa ^{\left( 1\right) }+\mathcal{%
O}(k_{3}^{2}).%
\end{array}%
\label{j12brownian}
\end{equation}

\section{Poisson Conductance}

An interesting use of our formalism is to compare the result of Eq.~(\ref{j12brownian})
with the case of Poisson noise. For Poisson we have~\cite%
{poisson},%
\begin{equation}
\begin{array}{ccc}
<\tilde{\eta}(s_{1})\tilde{\eta}(s_{2})>_{c} & = & \frac{2\,\gamma \,T}{%
s_{1}+s_{2}}, \\
&  &  \\
<\tilde{\eta}(s_{1})\tilde{\eta}(s_{2})\tilde{\eta}(s_{3})\tilde{\eta}%
(s_{4})>_{c} & = & \frac{1}{\lambda }\frac{24\,\,\gamma ^{2}\,T^{2}}{%
s_{1}+s_{2}+s_{3}+s_{4}},%
\end{array}%
\end{equation}%
where the expression for the fourth order cumulant can be checked by
dimensional analysis and compared with that of reference~\cite{poisson}.

The combinations $<\tilde{\eta}_{S,D}\tilde{\eta}_{S,D}>$ need to be
reexamined. For the Poisson noise, they yield,%
\begin{equation}
\begin{array}{ccc}
<\tilde{\eta}_{D}(s_{1})\tilde{\eta}_{D}(s_{2})>_{c} & = & \frac{2\,\gamma
\,(T_{1}+T_{2})}{s_{1}+s_{2}}, \\
&  &  \\
<\tilde{\eta}_{S}(s_{1})\tilde{\eta}_{S}(s_{2})>_{c} & = & \frac{\gamma
\,(T_{1}+T_{2})}{2\,(s_{1}+s_{2})}, \\
&  &  \\
<\tilde{\eta}_{S}(s_{1})\tilde{\eta}_{D}(s_{2})>_{c} & = & \frac{\gamma
\,(T_{1}-T_{2})}{s_{1}+s_{2}}, \\
&  &  \\
<\tilde{\eta}_{S}(s_{1})\tilde{\eta}_{D}(s_{2})\tilde{\eta}_{D}(s_{3})\tilde{%
\eta}_{D}(s_{4})>_{c} & = & \frac{1}{\lambda }\frac{12\,\,\gamma
^{2}\,(T_{1}+T_{2})\,(T_{1}-T_{2})}{s_{1}+s_{2}+s_{3}+s_{4}}.%
\end{array}%
\end{equation}%
The zero-th order on $k_{3}$ is exactly the same as the Gaussian case. The
first order can be illustrative and we shall calculate it in the following.
The quadratic term reads,%
\begin{equation}
\begin{array}{ccc}
\overline{\left\langle j_{12}\right\rangle _{12}} & = & k_{1}\lim_{z%
\rightarrow 0}\,\lim_{\epsilon \rightarrow 0}\,\lim_{\alpha \rightarrow
0}\,\int_{-\infty }^{\infty }\frac{dq_{1}}{2\pi }\,\ldots \int_{-\infty
}^{\infty }\frac{dq_{5}}{2\pi }\,\,\frac{z\,(iq_{2}+\epsilon )}{%
z-(iq_{1}+iq_{2}+2\epsilon )}\frac{2\,k_{3}}{R^{\prime }(iq_{1}+\epsilon )}%
\frac{\left\langle \tilde{r}_{D}(iq_{3}+\alpha )\,\tilde{r}%
_{D}(iq_{4}+\alpha )\,\tilde{r}_{D}(iq_{5}+\alpha )\,\tilde{r}%
_{S}(iq_{2}+\epsilon )\right\rangle _{c}}{iq_{1}+\epsilon
-(iq_{3}+iq_{4}+iq_{5}+3\alpha )} \\
&  &  \\
& = & 24\,\lambda \,\gamma ^{2}\,k_{1}\,k_{3}\lim_{z\rightarrow
0}\,\lim_{\epsilon \rightarrow 0}\,\lim_{\alpha \rightarrow
0}\,\int_{-\infty }^{\infty }\frac{dq_{1}}{2\pi }\,\ldots \int_{-\infty
}^{\infty }\frac{dq_{5}}{2\pi }\frac{z}{z-(iq_{1}+iq_{2}+2\epsilon )}\,\frac{%
(iq_{2}+\epsilon )}{iq_{1}+\epsilon -(iq_{3}+iq_{4}+iq_{5}+3\alpha )}\times
\\
&  &  \\
&  & \times (iq_{2}+\epsilon )\,\frac{(T_{1}+T_{2})\,(T_{1}-T_{2})}{%
(iq_{3}+iq_{4}+2\alpha +iq_{1}+iq_{2}+2\epsilon )} \\
&  &  \\
& = & -18\,{\gamma }^{2}\frac{k_{1}\,k_{{3}}}{\lambda }{\frac{\left( 8\,{k}%
^{2}{\gamma }^{2}m+20\,{\gamma }^{2}k\,m\,k_{1}+6\,k\,{\gamma }^{4}-9\,{%
\gamma }^{2}m\,{k}_{1}^{2}-4\,k\,{m}^{2}{k}_{1}^{2}-9\,{m}^{2}{k}%
_{1}^{3}\right) \left( {T_{{1}}}^{2}-{T_{{2}}}^{2}\right) }{\left[ {\gamma }%
^{2}\left( k+k_{1}\right) +m\,{k}^{2}\right] \left[ 6\,{\gamma }^{2}\left(
2\,k+3\,k_{1}\right) +16\,{k}^{2}m+72\,mk\,k_{1}+81\,m\,{k}_{1}^{2}\right]
\left( 3\,{\gamma }^{4}+4\,{\gamma }^{2}k\,m+4\,{\gamma }^{2}m\,k_{1}+{m}^{2}%
{k}_{1}^{2}\right) }}%
\end{array}%
\end{equation}%
The quartic term reads,%
\begin{equation}
\begin{array}{ccc}
\overline{\left\langle j_{12}\right\rangle _{14}} & = & -k_{3}\,\lim_{z%
\rightarrow 0}\,\lim_{\epsilon \rightarrow 0}\,\int_{-\infty }^{\infty }%
\frac{dq_{1}}{2\pi }\,\ldots \,\int_{-\infty }^{\infty }\frac{dq_{4}}{2\pi }%
\frac{z\,(iq_{2}+\epsilon )}{z-(iq_{1}+iq_{2}+iq_{3}+iq_{4}+2\epsilon )}%
\,\left\langle \tilde{r}_{D}(iq_{1}+\epsilon )\,\tilde{r}_{S}(iq_{2}+%
\epsilon )\,\tilde{r}_{D}(iq_{3}+\epsilon )\,\tilde{r}_{D}(iq_{4}+\epsilon
)\right\rangle _{c} \\
&  &  \\
& = & -12\,\lambda _{0}\,\gamma ^{2}\,k_{3}\,\lim_{z\rightarrow
0}\,\lim_{\epsilon \rightarrow 0}\int_{-\infty }^{\infty }\frac{dq_{1}}{2\pi
}\,\ldots \,\int_{-\infty }^{\infty }\frac{dq_{4}}{2\pi }\frac{z}{%
z-(iq_{1}+iq_{2}+iq_{3}+iq_{4}+2\epsilon )}\frac{1}{R^{\prime
}(iq_{1}+\epsilon )\,R^{\prime }(iq_{3}+\alpha )\,R^{\prime
}(iq_{4}+\epsilon )\,R^{\prime \prime }(iq_{2}+\epsilon )} \\
&  &  \\
&  & \times \frac{1}{R^{\prime }(iq_{1}+\epsilon )\,R^{\prime
}(iq_{3}+\alpha )\,R^{\prime }(iq_{4}+\alpha )\,R^{\prime }(iq_{5}+\alpha
)\,R^{\prime \prime }(iq_{2}+\epsilon )}\frac{(T_{1}+T_{2})\,(T_{1}-T_{2})}{%
(iq_{3}+iq_{4}+iq_{5}+3\alpha +iq_{2}+\epsilon )} \\
&  &  \\
& = & -18\,{\gamma }^{2}\frac{k_{1}\,k_{{3}}}{\lambda }{\frac{\left( 8\,{k}%
^{2}{\gamma }^{2}m+20\,{\gamma }^{2}k\,m\,k_{1}+6\,k\,{\gamma }^{4}-9\,{%
\gamma }^{2}m\,{k}_{1}^{2}-4\,k\,{m}^{2}{k}_{1}^{2}-9\,{m}^{2}{k}%
_{1}^{3}\right) \left( {T_{{1}}}^{2}-{T_{{2}}}^{2}\right) }{\left[ {\gamma }%
^{2}\left( k+k_{1}\right) +m\,{k}^{2}\right] \left[ 6\,{\gamma }^{2}\left(
2\,k+3\,k_{1}\right) +16\,{k}^{2}m+72\,mk\,k_{1}+81\,m\,{k}_{1}^{2}\right]
\left( 3\,{\gamma }^{4}+4\,{\gamma }^{2}k\,m+4\,{\gamma }^{2}m\,k_{1}+{m}^{2}%
{k}_{1}^{2}\right) }.}%
\end{array}%
\end{equation}

The total Poisson conductivity at order 1 on $k_{3}$ reads
\begin{equation*}
\begin{array}{ccc}
\kappa _{Poisson} & = & \frac{1}{2}{\frac{\gamma \,k^{2}}{{\gamma }^{2}(k_{{l%
}}+k)+{k}^{2}m}}+\frac{3}{2}\,{\frac{{\gamma }^{3}k_{1}\,k_{{3}}\left(
2\,k+k_{1}\right) \left( {T_{{1}}+T_{{2}}}\right) }{\left( k+2\,k_{1}\right) %
\left[ m{k}^{2}+{\gamma }^{2}\left( k+k_{1}\right) \right] ^{2}}+} \\
&  &  \\
&  & {+}54\,{\gamma }^{4}\frac{k_{1}\,k_{{3}}}{\lambda }\,\frac{\mathcal{N}}{%
\mathcal{D}}\left( T_{{1}}+T_{{2}}\right) +O\left( k_{3}^{3}\right)  \\
&  &  \\
& = & \kappa ^{\left( 0\right) }+\kappa ^{\left( 1\right)}+\kappa ^{\left( s\right) }+\mathcal{O}\left( k_{3}^{3}\right)
\end{array}%
,
\end{equation*}
where
\begin{equation}
\begin{array}{ccc}
\mathcal{N} & = & 5\,k\,{\gamma }^{2}+3\,{\gamma }^{2}k_{1}+3\,m\,{k}%
_{1}^{2}+4\,{k}^{2}m+11\,m\,k\,k_{1} \\
&  &  \\
\mathcal{D} & = & \left[ {\gamma }^{2}\left( k\,+k_{1}\right) +m{k}_{1}^{2}%
\right] \left[ 6\,{\gamma }^{2}\left( 2\,k+3\,k_{1}\right) +16\,{k}%
^{2}\,m+72\,m\,k\,k_{1}+81\,m\,{k}^{2}\right] \times  \\
&  & \times \left[ 3\,{\gamma }^{4}+4{\gamma }^{2}m\,\left( k\,+k_{1}\right)
+{m}^{2}{k}_{1}^{2}\right] .%
\end{array}%
\end{equation}

\end{widetext}

\end{document}